# Enhancing and Mapping Thermal Boundary Conductance across Bonded Si-SiC Interface

Rulei Guo[1,#], Bin Xu[1,3,#], Fengwen Mu[2,*], Junichiro Shiomi[1,3,*]

[1] Department of Mechanical Engineering, The University of Tokyo, Bunkyo, Tokyo 113-8656, Japan
[2] Institute of Microelectronics, Chinese Academy of Sciences, Beijing 100029, China.
[3] Institute of Engineering Innovation, The University of Tokyo, Bunkyo, Tokyo 113-8656, Japan
# These authors contributed equally to this work.
* Corresponding authors.
E-mail addresses:     mufengwen@isaber-s.com (F. Mu)
                     shiomi@photon.t.u-tokyo.ac.jp (J. Shiomi)

## Abstract

SiC is a promising substrate for Si-on-insulator (SOI) wafers for efficient thermal management owing to its high thermal conductivity and large bandgap. However, fabricating a Si device layer on a SiC substrate with a high and uniform thermal-boundary conductance (TBC) at the wafer scale is challenging. In this study, a 4-inch Si-on-SiC wafer was fabricated using a room-temperature surface-activated bonding method, and the TBC was measured using the time-domain thermoreflectance (TDTR) method. The obtained TBC was 109 MW/m$^2$K in the as-bonded sample, improving to 293 MW/m$^2$K after annealing at 750 °C, representing a 78% increase compared to previously reported values for a Si–SiC interface formed by bonding methods. Such enhancement is attributed to the absence of an oxide layer at the interface. Furthermore, we assessed the actual spatial distribution of the TBC in the SOI system by combining the TDTR mapping with a mathematical model to remove the influence of random errors in the experiment. The spatial distributions before and after annealing were 7% and 17%, respectively. Such variation highlights the need to consider the TBC distribution when designing thermal management systems.

# 1. Introduction

Si-on-insulator (SOI) wafers are widely used in the industry because of their low parasitic capacitance, catering to various applications, such as large-scale integration, photonics, microelectromechanical system, and radio-frequency chips[1]. Common SOI wafers consist of a Si substrate, buried $SiO_2$ insulator layer, and thin top-Si device layer, where electronic devices are built[2]. However, the low thermal conductivity of the $SiO_2$ layer impedes efficient heat dissipation from the device layer to the substrate, thereby increasing the temperature[3,4]. Such issue is critical for specialized applications, such as space technology and high-power-density scenarios, which negatively impact device performance, reliability, and lifetime[5]. To address this, replacing the oxide layer with materials with high thermal conductivity is a straightforward approach with SiC as a promising candidate material[4,6]. However, heat dissipation from the Si device layer to the SiC substrate is limited by their thermal-boundary resistance (TBR), that is, the inverse of the thermal-boundary conductance (TBC).

Several experimental and simulation studies have investigated the TBC at the Si–SiC interface. Xu et al.[7] used a nonequilibrium molecular dynamics (MD) simulation of a Si/4H–SiC interface to significantly modulate the TBC to 1000 $MW/m^2K$ by confining nanopatterns with nanothicknesses. Cheng et al.[8] measured a TBC of 620 $MW/m^2K$ for an epitaxially grown 3C–SiC/Si interface using the time-domain thermoreflectance (TDTR) method. Such value is among the highest obtained for semiconductor interfaces. However, nanopatterning and epitaxial growth are expensive and time-consuming methods that are impractical for large-scale thermal management.

Compared to other methods, the wafer-bonding method is an improved scalable and low-cost choice to achieve a wafer-scale uniform high TBC. Field et al.[5] utilized a hydrophilic bonding process to fabricate Si-on-4H–SiC samples. A TBC of 167 $MW/m^2K$ was obtained, which was much lower than that obtained using the epitaxial growth method. Moreover, high-resolution scanning transmission electron microscopy (HRSTEM) revealed an uneven $SiO_2$ layer at the interface, which can induce strong spatial nonuniformity in the TBC. Nonetheless, the evaluation of wafer-scale TBC inhomogeneity remains a challenge. In particular, the TBC spatial distribution in an SOI system is yet to be reported.

In this study, a 4-inch Si-on-4H–SiC bonded wafer was fabricated using the surface-activation bonding (SAB) technique, followed by annealing. The microstructure and composition of the interface were characterized by HRSTEM and energy-dispersive X-ray spectroscopy (EDS). Raman spectroscopy and HRSTEM were used to investigate the stress/strain of the Si device layer on the SiC substrate. The TBC and its spatial distribution across the bonded interfaces were evaluated using dual-modulation-frequency TDTR mapping with additional mathematical processing.

## 2. Results and Discussion

### 2.1 Sample preparation and structure characterization

A 4-inch SOI–SiC wafer was fabricated using the SAB method. As shown in Fig. 1a, Ar fast atom bombardment (Ar-FAB) was used to clean and activate the surfaces of the SOI and SiC. The two wafers were then pressed together under a pressure of 5 MPa to form a wafer-bonding interface. All processes were performed at room temperature under an ultrahigh vacuum. The structure of the as-bonded SOI–SiC wafer is shown in Fig. 1b. Subsequently, the Si substrate and buried oxide layer of the SOI wafer were removed by wet etching, and the sample was cut to obtain Si-on-SiC chips for further investigation, as shown in Fig. 1c. The samples were then observed under an infrared (IR) microscope. As depicted in Fig. S1, few bubbles or defects were observed at the bonding interface, indicating the high quality and macroscale uniformity of the Si/SiC interface. In addition, annealing at 750 °C was performed in an Ar environment for 30 min to further improve the crystallinity of the interfacial layer and enhance the TBC of the bonded interface.

The microstructure and composition of the Si-SiC interfaces before and after annealing were analyzed by HRSTEM and EDS (Fig. 1d). The high-angle annular dark-field (HAADF) image depicted two amorphous layers at the interface before annealing: an amorphous Si (a-Si) layer with a thickness of 3.8 nm and an amorphous SiC (a-SiC) layer with a thickness of 2.2 nm. These amorphous layers were formed by the Ar-FAB during the surface cleaning and activation steps. After annealing, the a-Si layer was crystallized, whereas the a-SiC layer remained with a thickness of 2.3 nm because the crystallization temperature of SiC is more than 800 °C, whereas that of Si is approximately 600 °C[9,10]. In the EDS image, Ar exists randomly at the a-Si and a-SiC layers, originating from the implantation of the Ar-FAB[11]. Oxygen and other elements were not noted at the interface owing to the cleaning process using Ar-FAB and high-vacuum environment. In addition, no apparent variation in the elemental distribution was observed, indicating that the interfacial structure variation was dominated by crystallization rather than elemental diffusion. We did not perform annealing at temperatures above 800 °C to avoid the full crystallization of the a-Si and a-SiC layers and alleviate the thermal stress and its potential impact on the electrical, optical, and thermal properties of the Si layer[12–14] because the a-SiC layer can function as a buffer layer to release the thermal stress originating from the mismatch of space group and lattice constant between Si (Fd$\bar{3}$m, a = 5.431 Å) and 4H–SiC (P6$_3$mc, a = b = 3.08 Å, c = 10.07 Å)[15].

Fig. 2a shows a high-resolution HAADF image of the interface after annealing to verify the lattice distortion near the interface. We subtracted the atomic position from the HAADF image and obtained the lattice constant to evaluate the strain, as shown in Fig. 2b. The dumbbell pattern formed by the adjacent Si atoms ensured the accuracy of the atom-position measurement. No substantial variations in the lattice constant were noted at different locations along the direction normal to the interface. In particular, the average lattice constant of 5.431 ± 0.003 Å, was similar to the reference value for bulk

Si (5.431 Å), as shown in Fig. 2c[16]. This suggests that the minimal strain near the interface.

Raman mapping measurements were performed to evaluate the stress distribution of the Si layer on a larger scale, as displayed in Fig. 2d. The optical phonon-mode Si peak at 520.6 cm$^{-1}$ red-shifted under tensile stress and blue-shifted under compressive stress. Assuming a uniaxial stress in the device layer, the Raman shift $\Delta\omega = \omega_0 - \omega$ (cm$^{-1}$) and stress exhibits a simple linear relationship as $\sigma = 434\Delta\omega$ (MPa)[17]. As illustrated in Fig. 2e, the stress of the Si layer is 18.5 ± 8.7 (tensile stress) and −74.3 ± 10.8 MPa (compressive stress) before and after annealing, respectively. The compressive stress after annealing arises from the annealing process because the linear thermal expansion coefficient of Si (2.45 ± 0.04 × 10$^{-6}$ K$^{-1}$) is smaller than that of SiC (4.13 × 10$^{-6}$ for the a axis and 3.78 × 10$^{-6}$ K$^{-1}$ for the c axis). Using the Young's modulus of Si, the strain in the Si device layer was calculated to be less than 0.07%, which has a negligible influence on the electrical and thermal properties of Si[12,13]. This result clarifies the importance of the a-SiC buffer layer in suppressing the thermal stress, as discussed above[15].

**2.2 Thermal boundary conductance at bonded interface**

TDTR was used to measure the TBC of the Si-SiC interface in the SAB-bonded wafer[18]. Because of the long mean free path of Si, the thermal conductivity of the thin Si device layer differed from its bulk value. Therefore, this measurement has three unknown thermal properties, namely the TBC of the interface between aluminum and Si (TBC$_{Al/Si}$), thermal conductivity of Si (TC$_{Si}$), and TBC of the interface between Si and the SiC substrate (TBC$_{Si/SiC}$). As fitting all these unknown parameters concurrently may lead to large uncertainty, we adopted a strategy using a dual-modulation-frequency measurement. As shown in Fig. S2a, the modulation frequency could be used to tune the thermal penetration depth, reflecting the probing depth of the sample from the surface. Therefore, the sensitivity of the individual parameters can be adjusted, denoting the extent to which an unknown parameter affects the measured signal, as shown in Fig. S2b. With the modulation frequency of 11 MHz, the thermal penetration depth was smaller than the thickness of the Si layer, resulting in the negligible sensitivity of TBC$_{Si/SiC}$. With the modulation frequency of 1 MHz, the thermal penetration depth was greater than the thickness of the Si layer, which increased the sensitivity of the TBC$_{Si/SiC}$. Therefore, TBC$_{Al/Si}$ and TC$_{Si}$ were measured at the modulation frequency of 11 MHz and used to obtain TBC$_{Si/SiC}$ at the modulation frequency of 1 MHz.

We used TDTR mapping to evaluate the TBC distribution in the study area. Unlike the commonly used TDTR mapping analysis, which samples only one data point at a particular delay time for each spatial point[19,20], we measured the full delay time range for accuracy. Mapping measurement was performed in a 100 μm × 100 μm area with 10 × 10 steps and repeated five times at the same area. The TDTR measurement signal and corresponding fitting curve used to obtain TBC$_{Si/SiC}$ before and after annealing are shown in Fig. 3a and 3b, respectively. The TBC$_{Si/SiC}$ was determined to be 109 MW/m$^2$K before annealing, increasing to 293 MW/m$^2$K, which is considerably larger than that of the previously reported SiC–Si sample fabricated through the hydrophobic wafer-bonding process[5]. The standard deviation of the TBC from the mapping measurement was 9.3 and 75.4 MW/m$^2$K before and after annealing, respectively.

The acoustic mismatch model (AMM) and diffuse mismatch model (DMM), which are widely used to predict the TBC between bulk crystal materials under the assumption of elastic scattering[21,22], were used to elucidate the thermal conduction across the bonding interface between Si and SiC. AMM assumes that phonons incident at an interface undergo specular transmission, whereas DMM assumes that the interface is completely disordered and that phonons lose their memory after reaching the interface[23]. As shown by dashed lines in Fig. 3c, the TBR at room temperature was predicted to be 8.7 and 2.4 m$^2$K/GW by DMM and AMM, respectively. This translates to 115.3 and 421.3 MW/m$^2$K in terms of TBC, respectively. The measured TBC before annealing closely aligned with the DMM results, indicating the diffuse scattering of the phonon transport at the interface. However, the measured TBC after annealing was notably larger than the DMM prediction and closer to the AMM prediction, implying the significant role of specular phonon scattering in this case. The change from diffuse to specular phonon scattering mechanisms can be attributed to variations in the amorphous interfacial layer at the interface, where the amorphous Si interlayer disappeared after annealing, thereby reducing the effective magnitude of disorder at the interface.

Thermal circuit calculation (TCC) provides an alternative approach to investigate the changes in the TBC before and after annealing[24]. The total TBR, which is the reciprocal of TBC, can be regarded as the sum of a series of thermal resistances using this method, as illustrated in Fig. 3c. Additional calculation details are presented in the Methods section. The total TBR were calculated to be 7.84 and 3.19 m$^2$K/GW, which is equivalent to 127.6 and 313.5 MW/m$^2$K before and after annealing, respectively, which agreed well with our measured results. This calculation suggests that crystallization of the amorphous Si layer is the main reason for the increase in TBC[4].

**2. 3 Spatial uniformity analysis of TBC**

The spatial uniformity of TBC is another critical issue for practical applications. The spatial distribution is typically obtained by directly calculating the standard deviation of TBC$_{Si/SiC}$ measured by the mapping method. The corresponding histograms of TBC$_{Si/SiC}$ before and after annealing are shown in Fig. 4a and 4b, respectively. The Gaussian distribution fit depicts a notable increase in the standard deviation from 9.3 MW/m$^2$K to 75.4 MW/m$^2$K after annealing. However, random errors in the TDTR measurement can also influence this standard deviation. In particular, for a large TBC$_{Si/SiC}$, minor random fluctuations stemming from the TDTR signal can profoundly affect the resulting TBC$_{Si/SiC}$, thereby altering the measured standard deviation of the spatial distribution. To address this issue, the measurement uncertainty caused by random errors in the TDTR measurement should be separated from the actual spatial distribution of the TBC. One feasible and straightforward method to realize this and obtain a precise TBC at each position is to increase the number of repeated measurements at individual locations, thereby ensuring the accuracy of the measured spatial distribution. Nevertheless, for a high TBC, attaining a desirable reduction in the measurement uncertainty requires a substantially large number of mapping measurements, which makes it not practical because mapping measurements are time-consuming (more details about the analysis of the measurement times can be found in the Supporting Information). As such, we present a statistical methodology that provides improved spatial distribution estimates with fewer measurements without knowing the precise value at each position. Two assumptions are made for this analysis: (1) the true spatial distribution of TBC follows normal distribution with an expected value of $\mu_{spatial}$ and standard deviation of $\sigma_{spatial}$; and (2) the difference between the measured TBC and its actual value with inherent random measurement error follows normal distributions with an expected value of 0 and standard deviation of $\sigma_{error}$. Subsequently, we established the following relationship:

$$\begin{cases} \mu_{spatial} = \mu_{measured} \\ \sigma_{spatial}^2 = \sigma_{measured}^2 - \sigma_{error}^2 \end{cases} \quad (1)$$

where $\mu_{measured}$ and $\sigma_{measured}$ are the average value and standard deviation of the multiple repeated mapping measurements, respectively; and $\sigma_{error}$ represents the average value of the standard deviation measured at each mapping position (more details regarding the mapping are provided in the Methods section and Supporting Information).

Based on the above model and five measurement times of an area, $\mu_{spatial}$, $\sigma_{error}$, and $\sigma_{spatial}$ were estimated to be 109, 11.1, and 7.5 MW/m²K, respectively, before annealing, and 293, 84.4, and 49 MW/m²K, respectively We further applied the uncertainty calculation method proposed by Yang et al.[23] to confirm the fluctuations in the TDTR theoretical curve. Although $\sigma_{error}$ changes substantially after annealing, the fluctuation in the TDTR signal is relatively constant. Such large difference primarily stems from the difference in the sensitivity of TBC$_{Si/SiC}$ under different values, as further discussed in the Supporting Information.

The coefficient of variation in the true spatial distribution of TBC ($\sigma_{spatial}/\mu_{spatial}$) obtained by TDTR mapping and mathematical analyses is approximately 7% and 17% before and after annealing, respectively, denoting that annealing considerably increases the TBC and spatial inhomogeneity. The practical impact of 17% variation depends on the actual application of interest. Nonetheless, the results highlight the importance of measuring the spatial distribution of TBC. Thus, the current TDTR mapping and uncertainty analysis facilitate a feasible approach to evaluate TBC at the wafer scale, which is valuable in ensuring the thermal performance and yield in the large-scale production of SOI wafers.

We plotted the relationship between TBC$_{Si/SiC}$ and thickness change of the amorphous interlayer, which is the difference from the average measurement thickness, as shown in Fig. 4c, to further discuss the origin of the spatial distribution of TBC$_{Si/SiC}$ ($\sigma_{spatial}$). The thermal conductivity of the amorphous layer was assumed to be 1.4 W/mK[25]. If $\sigma_{spatial}$ originates from the thickness change of the amorphous interlayer, the standard deviation of the thickness change before and after annealing should be 0.88 and 0.82 nm, respectively (dashed lines in Fig. 4c). However, the actual thickness change, $\sigma_d$, measured from HRSTEM image before and after annealing are 0.35 and 0.30 nm, respectively (shadow area in Fig. 4c). $\sigma_d$ is ascribed to the roughness of the interface between the crystal and amorphous parts, as shown in the inset. This roughness is mainly ascribed to the Ar-FAB process before annealing[26] and the crystallization process after annealing. As shown in the Fig. 4c, $\sigma_d$ cannot entirely explain $\sigma_{spatial}$, that is, other factors play an important role in determining $\sigma_{spatial}$. A possible factor for such is the uneven distribution of Ar, as shown in Fig. 1d[27].

**2.4 Heat-dissipation performance of the device**

Finite element method (FEM) was used for the simulation of the steady-state heat conduction using COMSOL to demonstrate the application potential of the Si-on-SiC bonding wafer and the impact of a high TBC in thermal management. In this simulation, pure Si, SOI, and Si-on-SiC wafers with different TBC values were compared. A steady-state Fourier heat conduction equation was employed in this simulation. The upper and lower surfaces of the wafers were set as adiabatic, whereas the bottom surface was maintained at room temperature. We added a heat source at the surface center of these wafers to simulate a hotspot. Subsequently, the maximum temperature rise was compared, as shown in Fig. 5. With the same hotspot power, the maximum temperature rise in the Si-on-SiC bonding wafer was approximately half that of the Si wafer and less than 1/8 that of the SOI wafer, demonstrating the promising application prospects of the Si-on-SiC wafers.

We compared the maximum temperature rise in the Si-on-SiC wafers with different TBC values. The maximum temperature increase was almost saturated when the TBC reached 200 MW/m²K. This finding suggests that the proposed Si-on-SiC wafer with a TBC of 293 MW/m²K should have a good performance for thermal management applications. Although the spatial distribution of the TBC broadened after annealing, it had a negligible influence on the heat dissipation performance in our simulation case because of the low sensitivity of the device temperature to the fluctuations of sufficiently high TBC. Notably, for actual industrial applications, the enlarged spatial distribution after annealing may result in local temperature fluctuations, which also prompts us to consider the spatial distribution of TBC.

## 3 Conclusions

In this study, a 4-inch Si-on-SiC wafer-bonded sample was fabricated using Ar-FAB as an alternative to SOI wafers for efficient thermal management. The Si-on-SiC sample was annealed to enhance its heat dissipation. The wafer-bonded sample demonstrated a high macroscale uniformity. Amorphous Si and SiC layers were observed before annealing, whereas only an amorphous SiC layer was present at the interface after annealing. We analyzed the stress/strain value and its distribution in the Si device layer. The low stress has negligible influence on the electrical and thermal properties of Si. Subsequently, the TBC of the bonding interface was measured using a dual-frequency TDTR method. The results showed a remarkable enhancement in the TBC from 109 MW/m$^2$K before annealing to 293 MW/m$^2$K after annealing. AMM, DMM, and thermal circuit model analyses were used to elucidate this enhancement. In addition, we combined the TDTR mapping method with a mathematical model to mitigate the influence of inevitable measurement noise and accurately quantify the spatial distribution of the TBC. Consequently, the TBC and its spatial distribution before and after annealing were revealed to be 7.5 MW/m$^2$K and 6.9% and 49 MW/m$^2$K and 16.7%, respectively. The quantitative analysis of the TBC spatial distribution can be ascribed to the roughness of the bonded interface and uneven Ar distribution. Therefore, this work proposes a promising candidate for thermal management materials and provides an advanced methodology for investigating TBC distribution.

# 4 Experimental section

## 4.1 Sample preparation

A 4-inch SOI wafer was bonded to a 4-inch 4H–sSiC wafer using the SAB method. The thickness of the device and buried oxide layers of the SOI wafer was approximately 2.4 and 2.0 μm, respectively. After cleaning and activating the surfaces of the SOI and SiC wafers by Ar-FAB, the two wafers were bonded at room temperature under a pressure of 5 MPa. All processes were performed in an ultra-high-vacuum environment to avoid oxidation or other contamination of the bonding interface. Subsequently, the Si substrate of the SOI was removed using a 25% KOH solution at 80 °C for several hours, whereas the $SiO_2$ layer of the SOI was removed using a 20% buffered hydrofluoric solution at room temperature for 30 min. Only the Si device layer remained on the SiC substrate.

## 4.2 IR microscopy

An IR microscope (SOM3355-IR, SII Nano Technology) was used to detect defects at the bonding interface. The IR microscope utilized a long-pass filter to illuminate light with a wavelength of 1300 nm, whereby the absorption depth for Si is more than 100 m[28], that is, the Si layer is almost transparent, facilitating defect observation at the bonding interface.

## 4.3 Microstructure analysis

The interfacial microstructure was characterized using HRSTEM (JEM-ARM200F Thermal FE STEM, JEOL Ltd.), and the composition was measured using EDS. During observation, the electron beam was aligned in the [011] direction of the Si crystal. The cross-section STEM samples were prepared with a focused ion beam (FIB)-scanning electron microscopy (SEM) system (XVision200TB, Hitachi High-Tech Corporation). A carbon layer was deposited at the beginning to protect the sample during the FIB process, and both sides of the sample were cleaned with Ar ions for 3 min after all other FIB processes to improve the surface quality and thereby the STEM image quality.

## 4.4 Raman spectroscopy

Raman spectroscopy (inVia confocal Raman microscope, Renishaw plc.) was used to characterize the stress in the device layer. Raman measurements were performed with a 532 nm laser, 1800 l/mm diffraction grating, 50× objective lens, and no polarization plate[17]. Raman mapping was performed to determine the stress distribution. Before mapping, the instrument was calibrated using a standard reference Si sample, and the focus condition was validated across the entire mapping area. After mapping, the baseline and effects caused by cosmic rays were removed.

## 4.5 Time-domain thermoreflectance

TDTR is a widely used thermal-property measurement method based on the pump-probe technology. A modulated pump beam periodically heats the surface, and a delayed probe beam detects surface temperature variations via thermoreflectance. The signal picked up by the photodiode and lock-in amplifier was fitted with an analytical heat conduction solution based on Fourier's law[29,30]. The wavelengths of the pump and probe beams were 400 and 800 nm, respectively. Both beams were focused on the sample surface by a 10× objective lens with the $1/e^2$ beam diameters of 15 and 10 μm, respectively. The Al transducer layer with a thickness of 80 nm was deposited on the sample surface using a vacuum evaporator. The thickness of the Al layer and thermal conductivity of SiC were measured on the reference samples using TDTR. The distribution of the thermal properties was measured using the TDTR mapping method. A three-axis motorized stage (KWC04015-C, Suruga Seiki Co., Ltd.) with a positioning accuracy of 1 μm was used to automatically change the measurement position. Prior to mapping measurements, the focus condition was validated for the entire mapping area. The mapping measurement was performed in a 100 μm × 100 μm area with 10 × 10 steps and repeated five times at the same area. The thermal penetration depth, $d_p$, was calculated as[25]

$$d_p = \sqrt{\frac{k}{\pi C_v f_{mod}}} \quad (2)$$

where $k$, $C_v$, and $f_{mod}$ represent the thermal conductivity, volumetric heat capacity, and modulation frequency, respectively. The TDTR sensitivity is defined as[31]

$$S_i = \frac{\partial \ln(-V_{in}/V_{out})}{\partial \ln(p_i)} \quad (3)$$

where $S_i$ is the sensitivity to parameter $i$, $V_{in}/V_{out}$ is the TDTR signal, and $p_i$ is the value of parameter $i$. A sensitivity of zero indicates that the parameter does not affect the measured signal.

The measured standard deviation of mapping parameter, $\sigma_{measured}$, was calculated as

$$\sigma_{measured}^2 = \frac{\sum_i \sum_{x,y}(p_{x,y,i}-\overline{p_i})^2}{n \times m} \quad (4)$$

where $p_{x,y,i}$ is the ith measured value at position $(x,y)$; $\overline{p_i}$ is the average value of the ith measurement over all positions; and $n$ and $m$ are the total number of measurement positions and number of repetitions, respectively. The standard deviation caused by measurement noise, $\sigma_{error}$, was calculated as

$$\sigma_{error}^2 = \frac{\sum_{x,y}\sum_i(p_{x,y,i}-\overline{p_{x,y}})^2}{n \times m} \quad (5)$$

where $\overline{p_{x,y}}$ is the average value at position $(x,y)$ over all the measurement times. Additional details are provided in Supporting Information.

The relationship between the noise level and TBC was estimated using the uncertainty analysis method proposed by Yang et al[32]. Noise was emulated by introducing uncertainty into the TDTR signal, while maintaining a consistent signal at each delay time.

**4.6 Phonon transmission models**

AMM and DMM calculations are widely used to predict the TBC between bulk crystal materials under the assumption of elastic scattering[33]. At the interface between material A and B, an incident phonon with frequency $\omega$ and polarization $j$ can either back-scatter or transmit. According to the Landauer formula, TBC, defined as $G$, can be predicted as follows[23]:.

$$G = \frac{1}{4}\sum_j \int_0^{\omega_A^{max}} \zeta_j^{A\to B} \hbar\omega v_{A,j} \cdot n_\perp D_{A,j}(\omega) \frac{\partial}{\partial T} f_0 \, d\omega \qquad (6)$$

where $\zeta$ is the phonon transmission coefficient, $\hbar$ is the reduced Planck constant, $v$ is the group velocity, $n_\perp$ is the unit normal vector of the interface, $D$ is the phonon density of state, $f_0$ is the Bose–Einstein distribution function, and subscript A and B indicates the two materials. In the AMM and DMM, the transmission coefficient is calculated as follows:

$$\zeta_{AMM}^{A\to B} = \frac{4\rho_A v_{A,j}\rho_B v_{B,j}}{(\rho_A v_{A,j}+\rho_B v_{B,j})^2} \qquad (7)$$

$$\zeta_{DMM}^{A\to B} = \frac{\sum_j D_{B,j} v_{B,j}}{\sum_j D_{A,j} v_{A,j}+\sum_j D_{B,j} v_{B,j}} \qquad (8)$$

where $\rho$ is the density, The phonon properties of Si and 4H–SiC were calculated using Phonopy, and the first-principles calculation was performed using VASP with the recommended projector augmented wave Perdew–Burke-Ernzerhof potential, as further discussed in the Supporting Information).

**4.7 Thermal circuit calculation**

Most thermal resistance values have been reported in the literature. The conductivity of the interface between crystalline and amorphous Si, $R_{c\text{-}Si/a\text{-}Si}$, was determined to be 1 m$^2$K/GW using MD[34]. The thermal conductivity of the amorphous SiC thin film (~2.5 nm) was calculated to be 1.4 W/mK by MD simulation [25]. The thermal conductivity of the amorphous Si thin film was inferred to be 1.02 W/mK from broadband frequency-domain thermoreflectance measurements of a-Si (500 nm)[35]. The conductivity of the interface between the amorphous and crystalline SiC, $R_{c\text{-}SiC/a\text{-}SiC}$, was investigated by MD to be 0.55 m$^2$K/GW[36]. However, some thermal resistance values have not yet been reported. Therefore, we assumed a common value. In particular, the TBR of the interface between amorphous Si and amorphous SiC, $R_{a\text{-}Si/a\text{-}SiC}$, and that between crystalline Si and amorphous SiC, $R_{c\text{-}Si/a\text{-}SiC}$, was assumed to be 1 m$^2$K/GW.

**4.8 FEM analysis**

A heat source was implanted at the wafer surface center with a heat flux of 300 W/mm$^2$ and diameter of 20 μm to simulate a hot spot. The maximum temperature increase was normalized to that of a pure Si wafer. The bottom temperatures of these wafers were fixed at room temperature, whereas the other surfaces were set to be adiabatic. An equal thickness of 400 μm was set for all these wafers. The thickness of the Si and SiO$_2$ layers of the SOI wafer were 2.4 and 2.0 μm, respectively. The thickness of the Si layer of the Si-on-SiC wafer was 2.4 μm.

## Acknowledgments

This research was funded in part by JSPS KAKENHI (grant nos. 22H04950 and 22K14189) and JST CREST (grant nos. JPMJCR1331 and JPMJCR21O2). We also thank Mari Morita from the Materials Advanced Research Infrastructure, University of Tokyo, for supporting the HRSTEM measurements.

# Figures

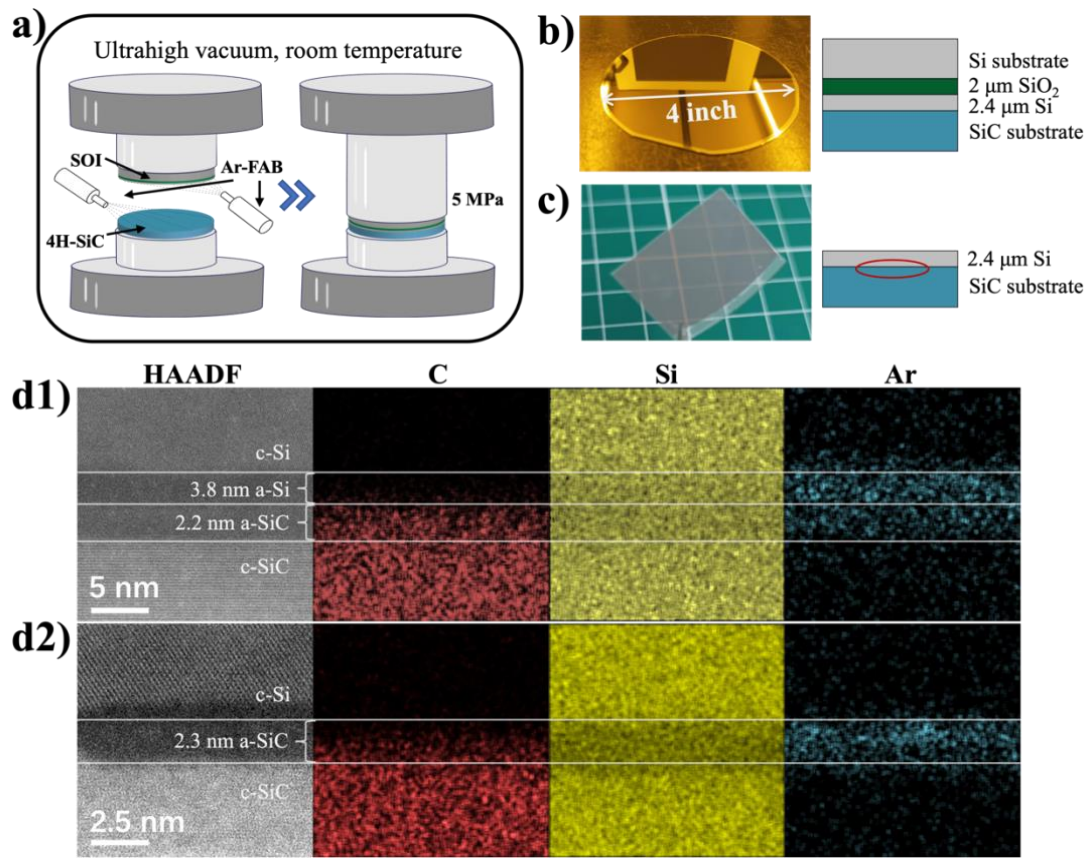

**Figure 1. Sample preparation and interfacial structure characterization.** A) Schematic of the SAB process. The optical image and the structure of b) the as-bonded 4-inch SOI–SiC wafer and c) an Si-on-SiC chip. The red circle marks the targeted Si–SiC interface. D) HRSTEM and EDS results at the SiC–Si bonding interface before annealing (d1) and after annealing (d2).

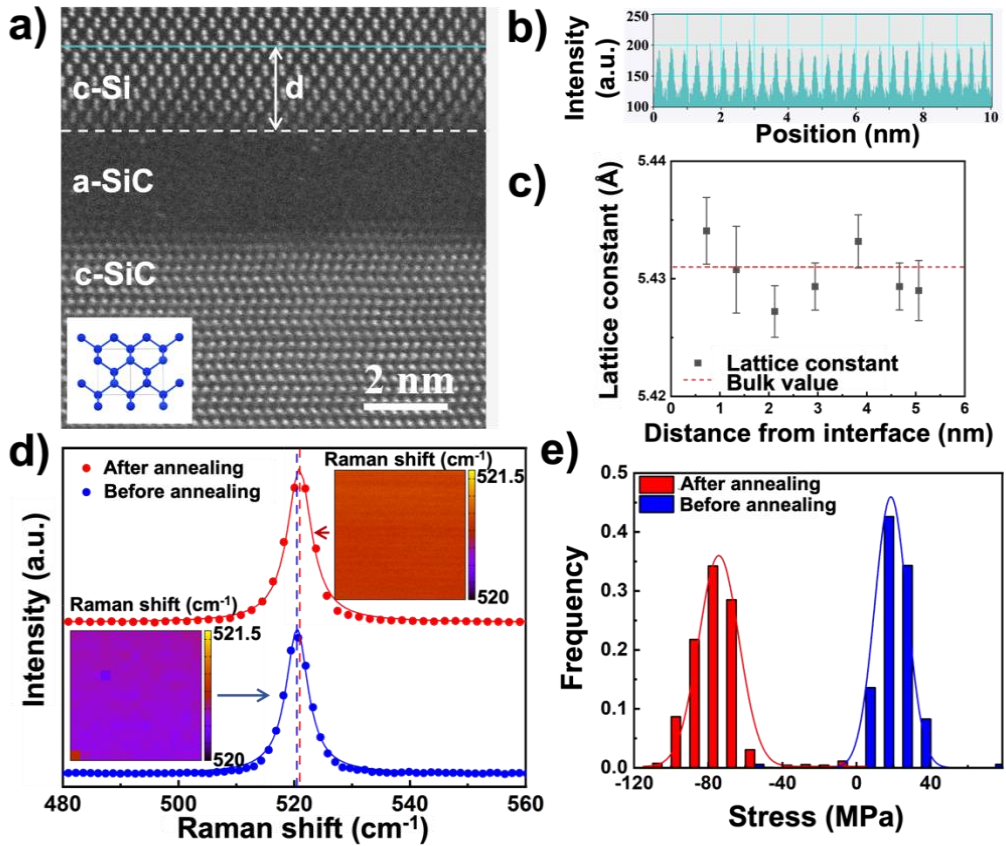

**Figure 2. Strain/stress characterization of the Si device layer.** A) HRSTEM image at the bonding interface after annealing. The inset shows the Si crystal structure observed from the [011] direction. B) Intensity of the HAADF image along the turquoise line in Fig. 2a. c) Lattice constant measured at different distances from the interface (marked as "d" in Fig. 2a). d) Raman spectrum. The solid line is the best fit obtained by the Lorentz distribution. The dashed lines indicate the peak position. The insets are the Raman peak-position maps of a 60 μm × 60 μm area. E) Stress distribution in the Si device layer calculated from the Raman peak-position mapping. The solid line is the normal-distribution fit.

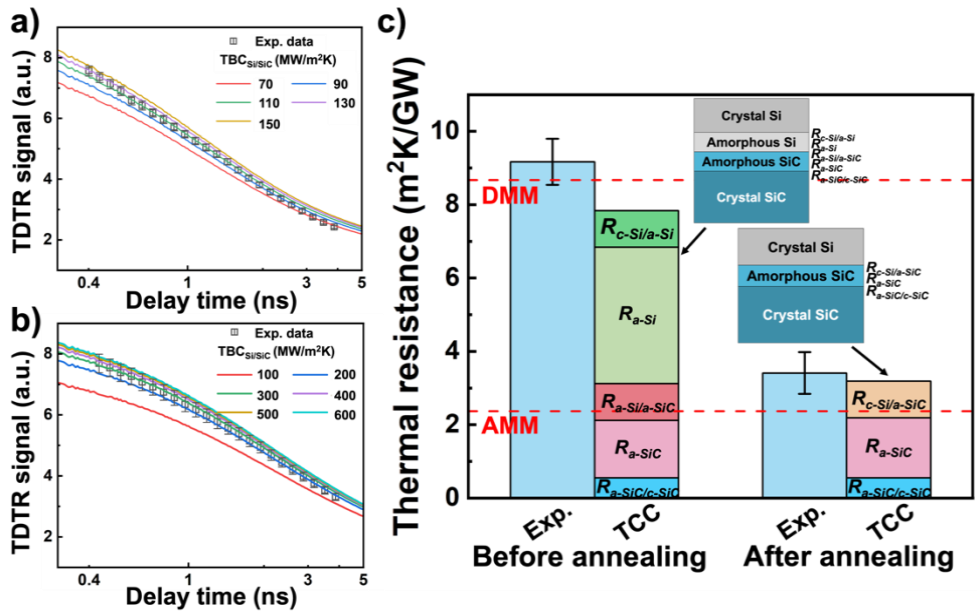

**Figure 3. TDTR measurement and analysis.** TDTR raw data and fitting lines before a) and after b) annealing. The error bars show the standard deviation of all mapping data. c) Comparison between TBR$_{Si/SiC}$ obtained by experimental methods, AMM, DMM, and thermal circuit calculation (TCC).

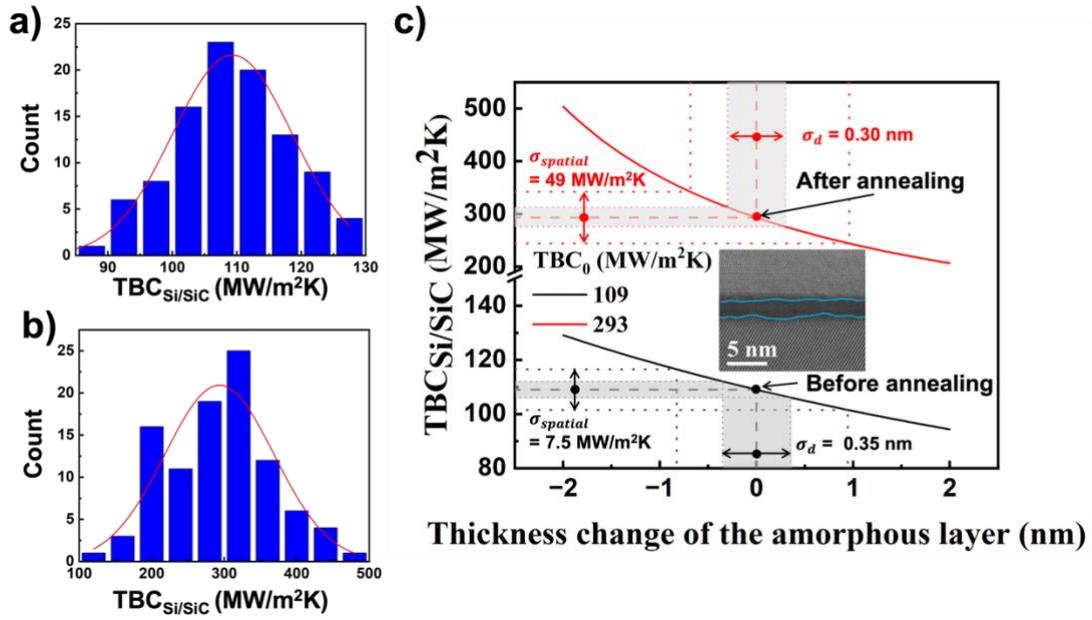

**Figure 4. Analysis of the spatial distribution of the TBC.** $TBC_{Si/SiC}$ mapping result averaged for five measurement before a) and after b) annealing. The red line indicates the best fit of the Gaussian distribution. c) Relationship between the thickness change of the amorphous layer and $TBC_{Si/SiC}$ for different $TBC_0$ (the value for the case without thickness change). The inset shows the roughness between the amorphous interlayer and crystal.

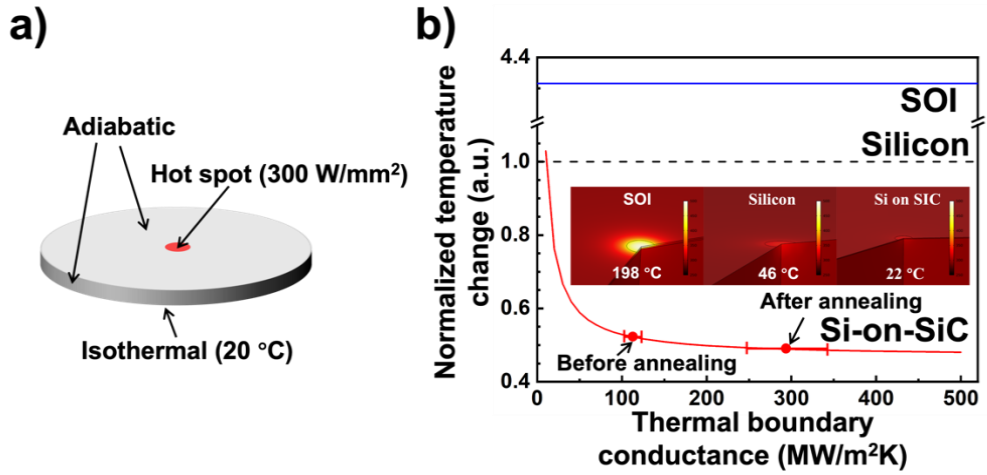

**Figure 5. FEM analysis.** a) Schematic of the FEM simulation setting. b) Maximum temperature change of the SOI, Si, and Si-on SiC wafers with different TBC values and the same hot-spot heating condition obtained by FEM. The maximum temperature change for Si was used for normalization. The average values and corresponding spatial distribution of our sample before and after annealing are depicted on the Si-on-SiC line.

# Supplementary Information

## Thermal Boundary Conductance Mapping across Bonded Heterogeneous SiC-Si Interface

Rulei Guo[#], Bin Xu[#], Fengwen Mu[*], Junichiro Shiomi[1,3,*]

# These authors contributed equally to this work.
*   Corresponding author.
E-mail addresses:    mufengwen@isaber-s.com (F. Mu)
                     shiomi@photon.t.u-tokyo.ac.jp (J. Shiomi)

**S.1 The infrared microscope image**

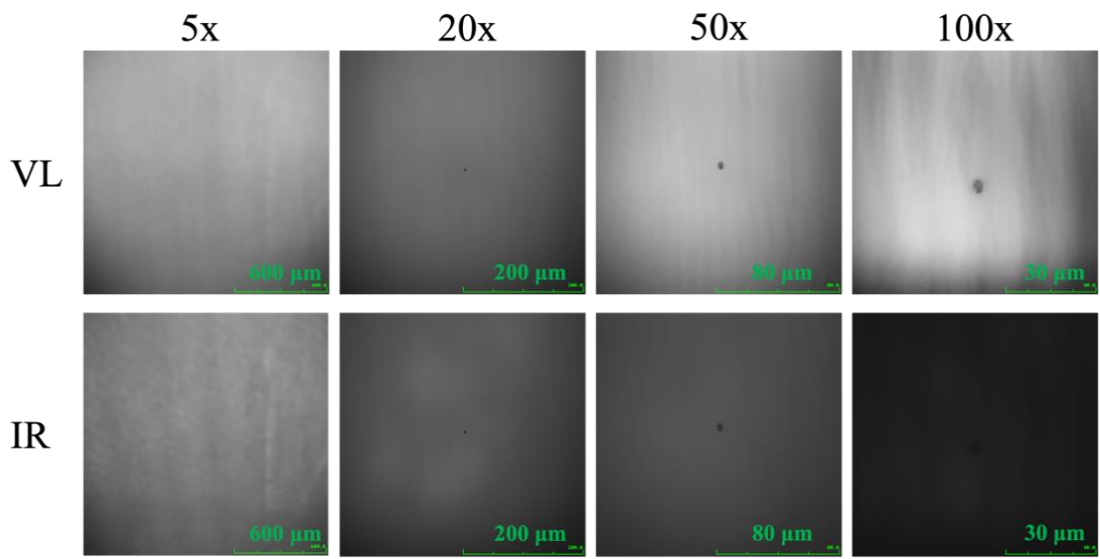

**Figure S1. The visual light (VL) and infrared (IR) images before annealing with magnifications 5-100X.** The black spot is a speck of dust on the surface which is used for focusing. Note: the shadow in the image may be caused by the dust or defect on light source of the microscope, as it remained constant while the sample was moving.

**S.2 Dual-modulation-frequency TDTR measurement**

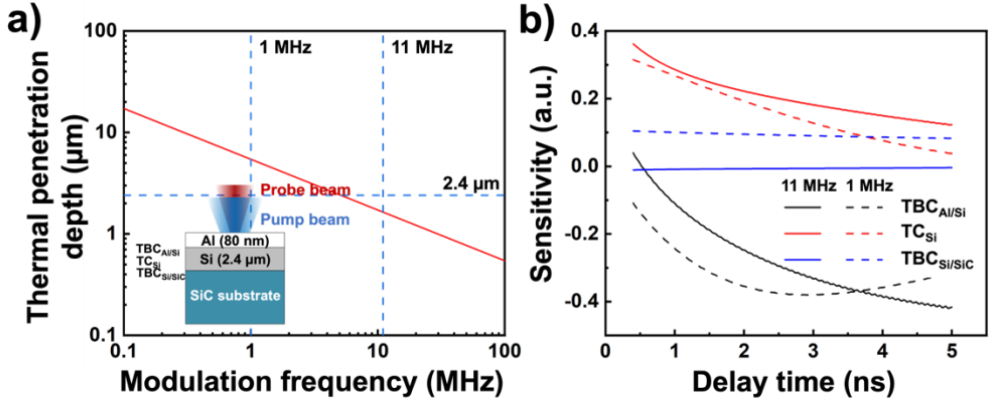

**Figure S2. Dual-modulation-frequency TDTR measurement.** a) Modulation-frequency-dependent thermal penetration depth for silicon. The inset shows the schematic of the TDTR method and the sample. b) The sensitivity of each fitting parameter with different modulation frequency.

### S.3 The mapping data analysis method

In this analysis, we make two key assumptions. Firstly, there are $p$ unknown parameters in the TDTR measurement, with their spatial distribution following the normal distribution $\mathcal{N}(X_{spatial}; \mu_{spatial}, Var_{spatial})$. Here, $X$, $\mu$ and $Var$ represent the random variable, the mean value and the variance matrix, respectively. Secondly, at each individual position, the difference between the measured result $X_{measured}$ and the actual value $X_{spatial}$ is assumed to follow the normal distribution $\mathcal{N}(X_{measured} - X_{spatial}; 0, Var_{error})$, attributable to the effect of the inherent random measurement noise. Then, the distribution of the measured value $P(X_{measured})$ can be calculated as below:

$$P(X_{measured}) = \int \mathcal{N}(X_{spatial}; \mu_{spatial}, Var_{spatial}) \times \mathcal{N}(X_{measured} - X_{spatial}; 0, Var_{error}) dX_{spatial}$$

$$= \mathcal{N}(\mu_{spatial}, Var_{spatial}) * \mathcal{N}(0, Var_{error})$$

$$= \mathcal{N}(X_{measured}; \mu_{spatial}, Var_{spatial} + Var_{error})$$

(S-1)

In scenarios where there is only one unknown parameter in the TDTR mapping measurement, we can establish the following relationship:

$$\begin{cases} \mu_{spatial} = \mu_{measured} \\ \sigma_{spatial} = \sqrt{\sigma_{measured}^2 - \sigma_{error}^2} \end{cases} \quad (S-2)$$

$\sigma$ is the standard deviation of the normal dsitribution.

To validate Equation S-2, we conducted a simulation of the TDTR mapping measuremnt. Firstly, a set of raw data (points number: 25 × 25) without noise was generated trough heat transfer model, the TBC of which follows normal distribution ($\mu = 250$ MW/m$^2$K, $\sigma = 30$ MW/m$^2$K). Secondly, we added random noise to these data to generate 5 sets of raw data with same noise level. Thirdly, we do the fitting process for these simulation raw data and calculate the spatial distribution utilizing Equation S-2. Finally, we compared the measured spatial distribution with the setted one. Two simulations with different noise level were performed and the result shows in Tabel S1. The difference between measured distribution and the setted one is less than 10%, indicating the validation of Equation S-2.

Tabel S1. The simulation of TDTR mapping

| Noise level (%) | $\mu_{spatial}$ (MW/m$^2$K) | $\sigma_{measured}$ (MW/m$^2$K) | $\sigma_{error}$ (MW/m$^2$K) | $\sigma_{spatial}$ (MW/m$^2$K) | Difference of $\sigma_{spatial}$ (%) |
|---|---|---|---|---|---|
| 1.5 | 252.4 | 31.1 | 10.5 | 29.3 | 2.3 |
| 3 | 253.8 | 38 | 20.1 | 32.3 | 8 |

## S.4 The thermal properties calculation for AMM and DMM model

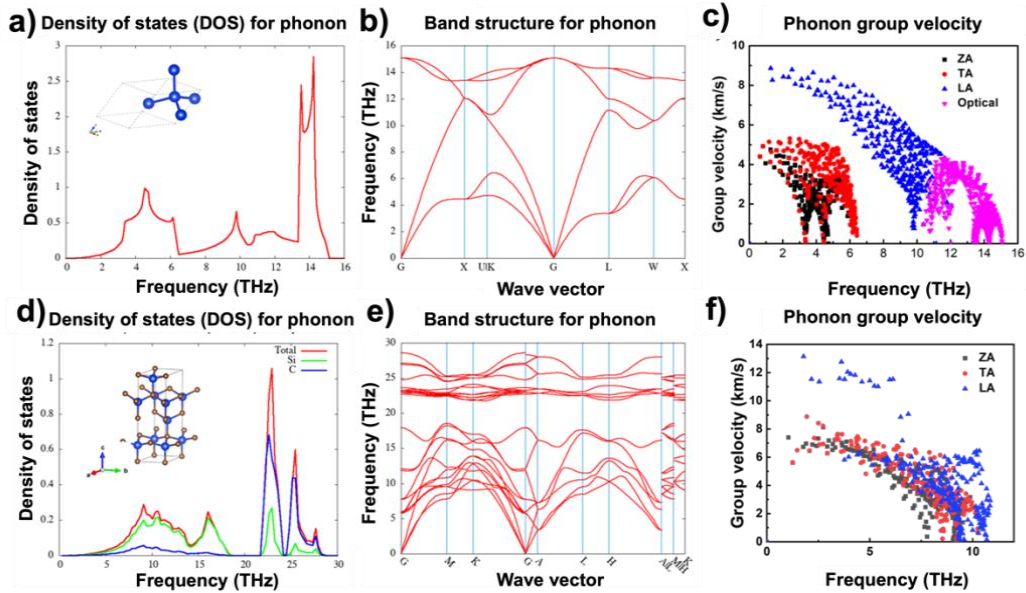

**Figure S3.1 The phonon property of silicon and 4H silicon carbide.** a)-c) is for silicon and the d)-f) for silicon carbide; a) and d) the phonon density of state; the inner picture is the lattice structure; b) and e) the phonon band structure; c) and f) the phonon group velocity versus frequency (only acoustic phonon of SiC was plotted here).

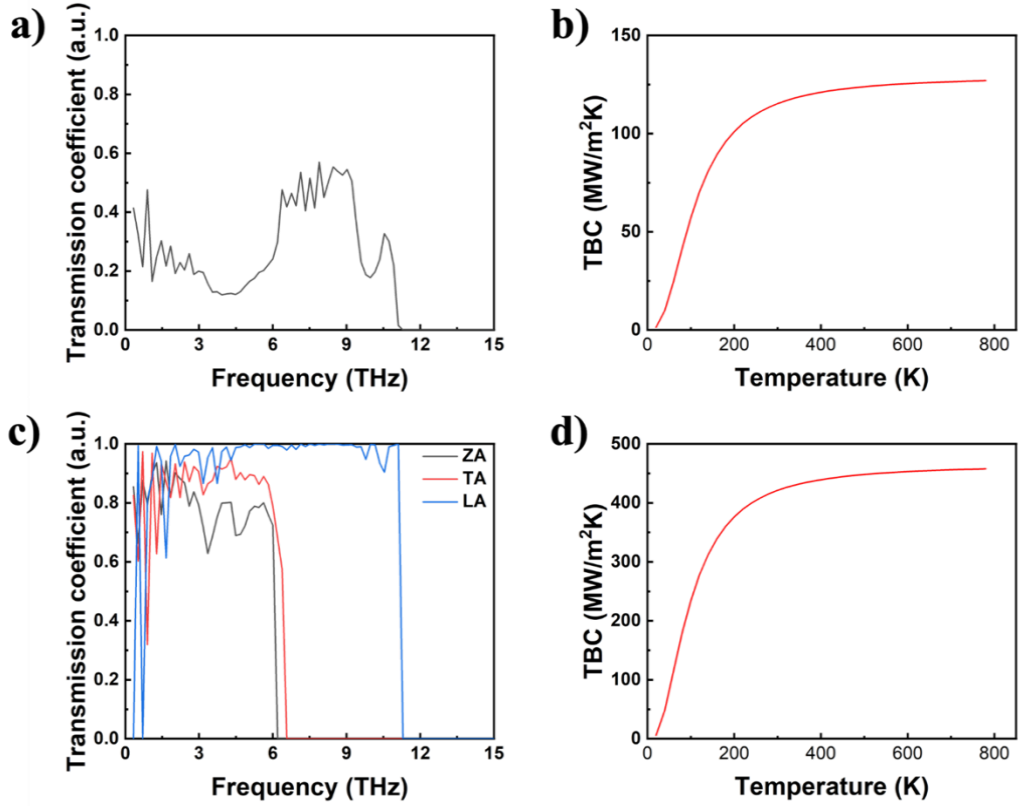

**Figure S3.2 The AMM and DMM model calculation for the interface between silicon and silicon carbide.** a) the phonon transmission coefficient of DMM; b) the temperature-dependent TBC predicted by DMM; c) the phonon transmission of AMM; d) the temperature-dependent TBC predicted by AMM.

**S.5 The relationship between the random error and the measurement times**

Due to the random fluctuation in the TDTR signal, the measured value of the TBC follows a Gaussian distribution with a standard deviation $\sigma_{error}$. For a n-time measurement, the mean value calculated from those measurements will have an associated standard error on the mean, $\bar{\sigma}_{error}$, given by[1]

$$\bar{\sigma}_{error} = \frac{\sigma_{error}}{\sqrt{n}} \qquad (S\text{-}3)$$

The relationship between the relative error $\bar{\sigma}_{error}/\mu$ and the measurement times before and after annealing is shown in Figure S4 ($\mu$ is the average value). We can see that a 5-time measurement is enough to reduce the relative uncertainty to 5% for the case before annealing, while this required time is 34 for the case after annealing.

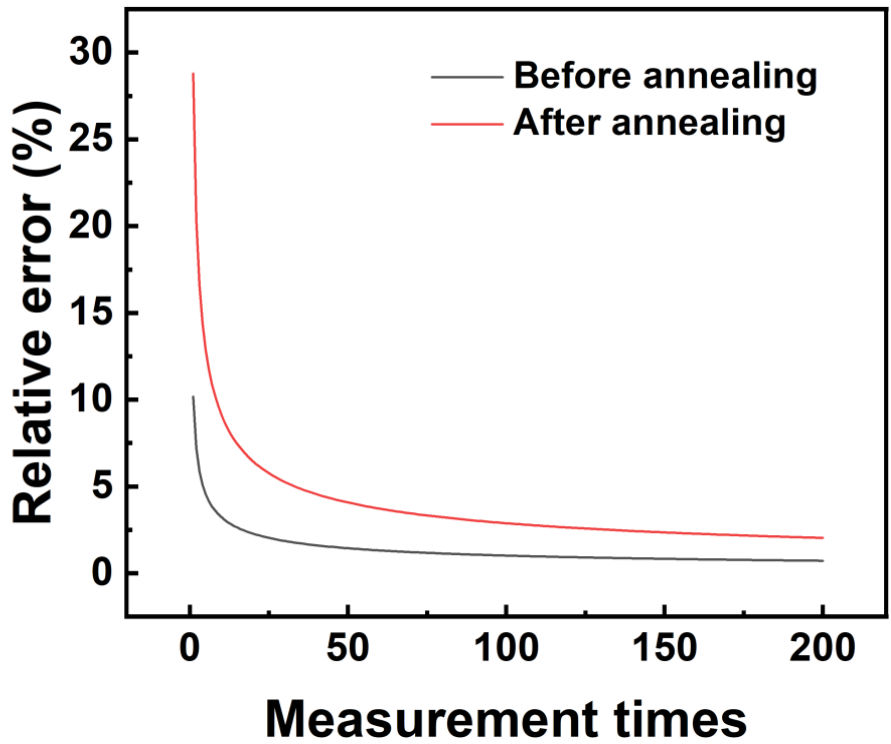

**Figure S4** The relationship between the relative error and the measurement times.

**S.6 The uncertainty calculation for determining the random fluctuation in TDTR signal**

We assume that the random fluctuation in TDTR signal follows a Gaussian distribution with a standard deviation $p \times \mu$. $\mu$ is the TDTR signal without noise. We call the $p$ noise level. As the $\sigma_{error}$ known, we found the corresponding $p$ through uncertainty analysis method proposed by Yang et al[2], as shown in Figure S5. Although the $\sigma_{error}$ changed by more than 7 times after annealing, the noise level just changes from 8.73% to 11.66% (~30% increasing in percentage). The noise level change may be caused by environment variation during the long mapping measurement.

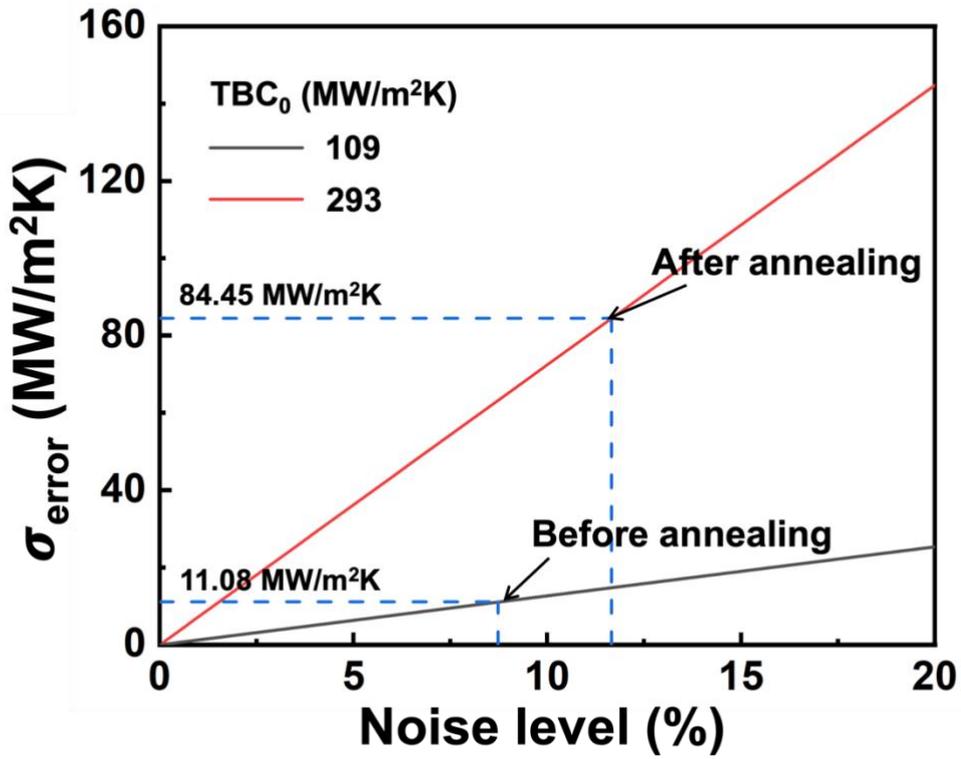

**Figure S5** The relationship between noise level and $\sigma_{error}$